\documentclass[11pt,a4paper]{article}
\usepackage[hyperref]{naaclhlt2019}
\usepackage{times}
\usepackage{latexsym}

\usepackage{url}

\usepackage{xspace}
\usepackage{xargs} 
\usepackage[table]{colortbl}

\usepackage{booktabs}

\usepackage{array}
\usepackage{graphicx}
\usepackage{subfig}

\usepackage{amsmath}
\definecolor{OliveGreen}{cmyk}{0.64,0,0.95,0.40}
\usepackage{comment}

\usepackage{bbding}
\usepackage{pifont}
\usepackage{wasysym}
\usepackage{amssymb}
\usepackage{amsthm}
\usepackage{mathtools}

\usepackage{commath}
\usepackage{pdflscape}
\usepackage{tabularx,array}

\usepackage{enumerate}
\usepackage{enumitem}

\usepackage{algpseudocode}

\usepackage{times}
\usepackage{fancyhdr}
\usepackage{multirow}

\usepackage{booktabs}
\usepackage{array,multirow,graphicx}

\usepackage{array}
\newcolumntype{P}[1]{>{\centering\arraybackslash}p{#1}}

\usepackage{xcolor}

\usepackage{setspace}

\usepackage[ruled,vlined]{algorithm2e}

\SetCommentSty{mycommfont}
\SetKwInput{KwInput}{Input}                
\SetKwInput{KwReturn}{return}

\usepackage{marginnote}

\newcommandx{\siva}[2][1=]{\todo[linecolor=red,backgroundcolor=red!10,bordercolor=red,#1]{SR: #2}\xspace}
\newcommand{\ignore}[1]{}

\usepackage{ragged2e,array,booktabs}
\usepackage{bbding}
\usepackage{pifont}
\usepackage{wasysym}
\usepackage{amssymb}
\usepackage{balance}
\usepackage{pbox}

\aclfinalcopy 

\title{Attention cannot be an Explanation}

\author{Arjun R Akula \\
  University of California, Los Angeles \\
  {\tt aakula@ucla.edu} \\\And
  Song-Chun Zhu \\
  University of California, Los Angeles \\
  {\tt sczhu@stat.ucla.edu} \\
  }

\date{}

\begin{document}
\maketitle
\begin{abstract}
    Attention based explanations (viz. saliency maps), by providing interpretability to black box models such as deep neural networks, are assumed to improve human trust and reliance in the underlying models. Recently, it has been shown that attention weights are frequently uncorrelated with gradient-based measures of feature importance. Motivated by this, we ask a follow-up question: ``Assuming that we only consider the tasks where attention weights correlate well with feature importance, how effective are these attention based explanations in increasing human trust and reliance in the underlying models?". In other words, can we use attention as an explanation? We perform extensive human study experiments that aim to qualitatively and quantitatively assess the degree to which attention based explanations are suitable in increasing human trust and reliance. Our experiment results show that attention cannot be used as an explanation.
\end{abstract}

\section{Introduction}
How to increase human trust and reliance in AI systems
is of great concern in a wide range of applications that depend on the machine's predictions and decisions such as medical diagnosis systems and self-driving cars~\cite{chancey2015role,gulshan2016development,lyons2017certifiable}. 
Previous studies have shown that trust is closely and positively correlated to the level of how much human users understand the system ({\bf understandability}) and how accurately they can predict the system's performance on a given task ({\bf predictability})~\cite{hoffman17explanation,hoffman2018metrics,miller2018explanation,hoffman2010metrics}. 
Despite an increasing amount of work on explainable AI (XAI) methods~\cite{smilkov2017smoothgrad,sundararajan2017axiomatic,akula2022discourse,zeiler2014visualizing,kim2014bayesian,zhang2018interpretable}, providing explanations that can increase understandability and predictability remains an important research problem~\cite{DBLP:journals/corr/abs-1903-02252,akula20words,akula2019visual,akula2021crossvqa,akula2021measuring,akula2021robust,akula2020cocox,r2019natural,pulijala2013web,gupta2012novel}.

Attention based explanations~\cite{bahdanau2014neural,smilkov2017smoothgrad,sundararajan2017axiomatic,zeiler2014visualizing}, by providing interpretability, are assumed to increase human trust and reliance in the underlying vision and NLP models. However, currently there is no work in the literature that formally evaluates this assumption. Quite recently, \cite{DBLP:journals/corr/abs-1902-10186} showed that attention mechanism do not correlate with gradient-based measures of feature importance and hence are not faithful. Motivated by this, we ask a follow-up question : ``If we consider only the tasks where attention is indeed faithful, how effective are these attention based explanations in increasing human trust and reliance in the underlying  models?"~\cite{akula2022effective,agarwal2018structured,akula2019natural,akula2015novel,palakurthi2015classification,agarwal2017automatic,dasgupta2014towards}. 

We perform extensive human study experiments to investigate the relationship between attention based explanations, and human trust and reliance. We also compare attention based explanations with non-attention based XAI frameworks using several XAI evaluation metrics defined in ~\cite{hoffman2018metrics}~\cite{akula2013novel,akula2018analyzing,akula2021mind,gupta2016desire,akula2019explainable,akula2021gaining,akula2019x,akula2020words,Akula2019DiscoursePI,Akula2022QuestionGF}.
Our findings show that attention based explanations are not effective to improve human trust, whereas the non-attention baselines significantly improve human trust. To the best of our knowledge, this is the first work to quantitatively and qualitatively compare all the state-of-the-art XAI frameworks and show that attention cannot be used as an explanation.



\section{Task and Datasets}
Authors in \cite{DBLP:journals/corr/abs-1902-10186} considered NLP tasks (e.g. text classification, natural language inference (NLI), and question answering) to show that attention mechanism is not faithful. However, our goal is to find whether if attention can improve human trust and reliance conditioned on the assumption that attention mechanism is faithful. Therefore, we chose an image classification task as there are several previous studies showing that attention weights correlate to the feature importance in image classification~\cite{selvaraju2017grad,zhang2018interpretable}. We experiment with Caltech-UCSD Birds dataset~\cite{wah2011caltech}. We consider a deep convolutional network VGG-16~\cite{simonyan2014very} for the classification task. 

\section{Attention and Non-Attention frameworks}

We experiment with the following five explainable AI (XAI) frameworks that provide attention based explanations. These frameworks are also called as Attribution techniques.\\
(1) \textit{Class Activation Mapping (CAM)} \cite{zhou2016learning}. CAM produces attention map as explanation, i.e. it highlights the important regions in the image for predicting a target output.\\
(2) \textit{Gradient-weighted Class Activation Mapping (Grad-CAM)} \cite{selvaraju2017grad}. Grad-CAM uses the gradients of target class flowing into the final convolutional layer to produce attention map as explanation.\\
(3) \textit{Local Intepretable Model-Agnostic Explanation (LIME)} \cite{ribeiro2016should}. LIME produces attention map as explanation, generated through superpixel based perturbation.\\
(4) \textit{Layer-wise Relevance Propagation (LRP)} \cite{bach2015pixel}. LRP generates attention map by propagating classification probability backward through the network and then calculates relevance scores for all pixels.\\
(5) SmoothGrad \cite{smilkov2017smoothgrad}. Smooth grad produces attention map as explanation by adding gaussian noise to the original image and then calculating gradients multiple times and averaging the results.

We use the following three non-attention XAI frameworks to compare against the above five attention based explanations:\\
(6) \textit{TCAV} \cite{kim2018interpretability}. In contrast to pixel-level explanations of attention maps, TCAV generates concept based explanations. It computes directional derivatives to produce estimates of how important a concept was for a CNN's prediction of a target class, e.g. how important is the concept of \texttt{stripedness} for predicting the zebra class.\\
(7) \textit{CEM} \cite{dhurandhar2018explanations}. CEM provides contrastive
explanations by identifying pertinent positives and pertinent negatives in the input image.\\
(8) \textit{Counterfactual Visual Explanations (CVE)} \cite{Goyal2019counterfactual}.  CVE provides counterfactual explanation describing what changes to the situation would have resulted in arriving at the alternative decision.

\section{Evaluation Metrics}
In our human study experiments, we use all the standard qualitative and quantitative XAI evaluation metrics~\cite{hoffman2018metrics} to evaluate the above attention and non-attention methods.\\ \\
\textbf{Quantitative Metrics}:\\
(1) \textit{Justified Positive and Negative Trust}. Trust is measured by evaluating the human's understanding of model's ($M$) decision making process. Let us consider that $M$ predicts images in the set $C$ correctly and makes incorrect decisions on the images in the set $W$. Justified positive trust (JPT) is given as the percentage of images in $C$ that the human subject felt $M$ will correctly predict. Similarly, Justified negative trust (JNT), also called as mis-trust, is given as the percentage of images in $W$ that the human subject felt $M$ will fail to predict correctly.\\ \\
(2) \textit{Reliance}. Reliance (Rc) captures the extent to which a human can accurately predict the model’s inference results without over- or under-estimation. In other words, Reliance is proportional to the sum of JPT and JNT.\\ \\
\textbf{Qualitative Metrics}:\\
(3) \textit{Explanation Satisfaction}. We  measure  human  subjects’ feeling of satisfaction at having achieved an understanding of the machine in terms of usefulness, sufficiency, appropriated detail, confidence, understandability, accuracy and consistency. We ask them to rate each of these metrics on a Likert scale of 0 to 9.



\section{Experiments}
We conduct extensive human subject experiments to assess the effectiveness of attention based explanations and non-attention based explanations, in helping human users understand the internal workings of the underlying model. We recruited 180 human subjects from our institution's Psychology subject pool~\footnote{These experiments were reviewed and approved by our institution's IRB.}. These subjects have no background on computer vision, deep learning and NLP. We applied between-subject design and randomly assigned each subject into one of the 9 groups: 8 groups for each of the XAI frameworks discussed in section 3 and one baseline group, called NO-X. Subjects in NO-X are not provided with any explanations for the model's predictions. 

Within each group, each subject will first go through a familiarization phase where the subjects become familiar with the underlying model, followed by a testing phase where we apply our evaluation metrics and assess their understanding in the underlying model. In the familiarization phase, the NO-X group will be shown several images and their predictions by the model, whereas each of the other 8 groups will be additionally shown explanations for the model predictions. We will make all our experiment data and results publicly available. 

\textbf{Quantitative Results}: We compares the quantitative metrics of all the 9 groups.
We find that trust (JPT, JNT) and reliance (Rc) values of attention based explanations do not differ significantly from the NO-X baseline. This clearly shows that attention based explanations are not effective to increasing human trust and reliance. On the other hand, non-attention baselines such as TCAV and CVE are significantly higher than the attention baselines. This could be attributed to the fact that concept based explanations in TCAV and counterfactual explanations in CVE are lucid and easily understandable by humans as they emphasize only on the critical concepts that are highly influential in the network’s decision making process.

\begin{figure}[t]
\centering
  \includegraphics[width=1\linewidth,height=0.88\linewidth]{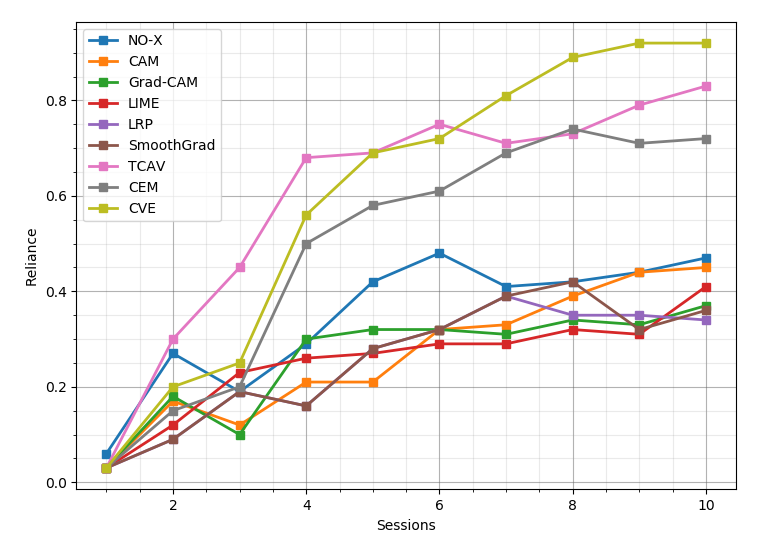}
  \caption{Gain in Reliance over time}~\label{fig:exp1}
  
\end{figure}

\textbf{Qualitative Results}: The average explanation satisfaction rates obtained from each of the 8 groups are shown in Table~\ref{tab2} (we do not include NO-X group as they are not shown any explanation). Human subjects in TCAV group found that explanations were highly useful, sufficient and detailed compared to the attention baselines ($p < 0.01$). As we can see, subjects in the attention baselines are clearly not satisfied with the explanations. It is interesting there aren't any significant differences across the 8 groups in terms of other explanation satisfaction measures: confidence, understandability, accuracy and consistency. We did an additional study with 4 subjects in each of the 8 groups to verify these four satisfaction measures and again found similar results. We leave this observation for future exploration.

\vspace{-4pt}
\begin {table*}[t]
\small
\begin{center}
\begin{tabular}{ | m{1.5cm} | m{1.5cm}| m{1.5cm}| m{1.6cm} | m{1.6cm}| m{2.2cm}| m{1.6cm} |} 
\hline
Fram. & Useful. & Suff. & Appro. & Confidence & Understand. & Accuracy\\ 
\hline
CAM  & 1.5 & 1.5 & 2.1 & 4.8 & 4.5 & 4.1  \\
\hline
Grad-CAM & 1.7 & 2.4 & 3.1 & 3.4 & 4.2 & 3.7 \\
\hline
LIME  & 2.1 & 1.1 & 1.5 & 2.8 & 3.1 & 2.9  \\
\hline
LRP  & 1.6 & 1.8 & 2.4 & 3.7 & 4.1 & 3.2 \\
\hline
Smooth.  & 2.1 &  2.3 & 3.4 & 4.6 & 3.1 & 2.9\\
\hline
TCAV  & 7.4 & 8.1 & 6.9 & 4.4 & 5.4 & 4.9 \\
\hline
CEM  & 6.1 & 5.95 & 6.5 & 4.1 & 5.3 & 5.2 \\
\hline
CVE  & 7.2 & 7.9 & 6.9 & 5.3 & 5.5 & 4.9 \\
\hline
\end{tabular}
\caption {Explanation Satisfaction using attention and non-attention baselines}
\label{tab2}
\end{center}
\end{table*}




\textbf{Competency Testing}:
We perform an additional experiment where we train three different CNNs namely; AlexNet, VGG-16 and ResNet-50. It may be noted that ResNet-50 is known to be more reliable than VGG-16, and VGG-16 is more reliable than AlexNet. We show the predictions and the explanations from each of the three networks to the subjects and ask them to rate the reliability (competency) of the models relative to each other on a Likert scale of 0 to 9 (we chose only those images for which all the three models made the same prediction as ground truth). The assumption here is that a good explanation helps the subject to distinguish between reliable model and unreliable model. We found that CVE significantly outperformed other baselines. TCAV and CEM performed better than NO-X baseline, whereas all the attention baselines did not perform any better than NO-X baseline.

\textbf{Gain in Reliance over Time}:
We hypothesized that, human trust and reliance in machine might improve over time. This is because, it can be harder for humans to fully understand the machine's underlying inference process in one single session. Therefore, we conduct an additional experiment with eight human subjects where the subjects' reliance is measured after every session. Note that each session consists of a familiarization phase followed by a testing phase. The results are shown in Figure~\ref{fig:exp1}. As we expected, subjects' reliance increased over time. All the three non-attention baselines significantly improved over time with CVE outperforming TCAV and CEM. Surprisingly, gain in Reliance for attention baselines is less than the NO-X baseline. These findings clearly suggest that attention cannot be considered as an explanation.





\section{Related Work}
\textbf{Attention based explanations}: Most prior work has focused on generating attention based explanations. Attention maps highlight pixels of the input image (saliency maps) that most caused the output classification. For example, gradient-based visualization methods~\cite{zhou2016learning,selvaraju2017grad} have been proposed to extract image regions responsible for the network output. The LIME method proposed by~\cite{ribeiro2016should} explains predictions of any classifier by approximating it locally with an interpretable model.  Influence measures~\cite{datta2015influence} have been used to identify the importance of features in affecting the classification outcome for individual data points. Other related work includes a visualization-based explanation framework for Naive Bayes classifiers~\cite{greiner2003explaining}, an interpretable  character-level language models for analyzing the predictions in RNNs~\cite{karpathy2015visualizing}, and an interactive visualization for facilitating analysis of RNN hidden states \cite{strobelt2016visual}.\\
\textbf{Non-Attention based explanations}: There are few attempts in the literature that go beyond the pixel-level explanations. For example, TCAV technique proposed by \cite{kim2018interpretability} aim at generating explanations based on high-level user defined concepts. Contrastive explanations are proposed by \cite{dhurandhar2018explanations} to identify minimal and sufficient features to justify the classification result. \cite{Goyal2019counterfactual} proposed counterfactual visual explanations that identify how the input could change such that the underlying vision system would make a different decision. More recently, explainable AI focuses on building models which are intrinsically interpretable~\cite{zhang2017interpretable,Akula2022QuestionGF}.


\section{Conclusions}
In this paper, motivated by the recent findings~\cite{DBLP:journals/corr/abs-1902-10186}, we pose a follow-up research question: how effective are the attention based explanations in improving human's understanding of underlying black box models. To answer this question, we conducted extensive human study experiments and evaluated the state-of-the-art attention and non-attention explainable AI (XAI) frameworks using qualitative and quantitative XAI metrics. Our results conclude that attention cannot be used as an explanation. 

\bibliographystyle{acl_natbib}
\bibliography{naaclhlt2019}

\end{document}